**Magnetic behavior of metallic kagome lattices, $Tb_3Ru_4Al_{12}$ and $Er_3Ru_4Al_{12}$**

**Sanjay Kumar Upadhyay, Kartik K Iyer, and E.V. Sampathkumaran**
*Tata Institute of Fundamental Research, Homi Bhabha Road, Colaba, Mumbai 400005, India*

Abstract

We report magnetic behavior of two intermetallics-based kagome lattices, $Tb_3Ru_4Al_{12}$ and $Er_3Ru_4Al_{12}$, crystallizing in the $Gd_3Ru_4Al_2$-type hexagonal crystal structure, by measurements in the range 1.8 – 300 K with bulk experimental techniques (ac and dc magnetization, heat-capacity and magnetoresistance). The main finding is that the Tb compound, known to order antiferromagnetically below ($T_N=$) 22 K, shows glassy characteristics at lower temperatures (<< 15 K), thus characterizing this compound as a re-entrant spin-glass. The data reveal that the glassy phase is quite complex and is of a cluster type. Since glassy behavior was not seen for the Gd analogue in the past literature, this finding on the Tb compound emphasizes that this kagome family could provide an opportunity to explore the role of higher order interactions (such as quadrupole) in bringing out magnetic frustration. Additional findings reported here for this compound are: (i) The plots of temperature dependence of magnetic susceptibility and electrical resistivity data in the range 12 – 20 K, just below $T_N$, are found to be hysteretic leading to a magnetic phase in this intermediate temperature range, mimicking disorder-broadened first-order magnetic phase transitions; (ii) features attributable to an interesting magnetic phase co-existence phenomenon in the isothermal magnetoresistance in zero field, after travelling across metamagnetic transition fields, are observed. With respect to the Er compound, we do not find any evidence for long-range magnetic ordering down to 2 K, but this compound appears to be on the verge of magnetic order at 2 K.

I.   Introduction

The field of frustrated magnetism arising from the geometry of the arrangement of magnetic ions in solids provides opportunities to explore new quantum states in condensed matter physics. In this respect.  kagome lattices have been presenting exotic situations. While kagome lattices are commonly known among oxides and many other inorganic as well as organic compounds,  kagome nets are not very commonly known among intermetallic compounds containing magnetic ions, barring some exceptions, e.g., rare-earth (R) compounds crystallizing in ZrNiAl-type structure [1], and $GdInCu_4$ [2]. It is therefore of interest to search for metallic kagome families (that is, which are mainly controlled by Ruderman-Kittel-Kasuya-Yosida indirect exchange interaction) and to study their magnetic behavior. It is also to be emphasized that the interplay between structural, orbital, charge and spin degrees of freedom  as well as crystallographic defects is at the heart of many solid state phenomena in the current literature. It is therefore of great interest to expand our knowledge in this direction in the field of geometrically frustrated magnetism.

In this respect, the hexagonal family of rare-earths and actinides of the type, $R_3Ru_4Al_{12}$, crystallizing in the space group P6$_3$/*mmc* [3, 4], has attracted some interest [5 – 12] in recent years. We have summarized  crystallographic features in our  article on the Gd compound [12]. The rare-earth ions, present in a layer,  form distorted kagome nets with larger and smaller triangles, with these triangles sharing vertices and with different edge lengths. The readers may see figures 1a and 1b for the structure produced using VESTS software  [13].  These R-layers are separated by puckered layers of $Ru_4Al_8$. Surprisingly, the prototype member [3] of this structure, $Gd_3Ru_4Al_{12}$, was studied for its magnetic behavior recently only [12], despite its discovery nearly quarter century ago. Various magnetic anomalies like multiple magnetic transitions  as a function of temperature and magnetic field (for instance in the single crystals of Tb and Dy members, Refs. 10, 11], intermediate-valence behavior in the case of Ce [7, 8] and  anomalous magnetoresistance loops have been reported in this family. We have noted that the investigations in this family, particularly for the Tb compound,  are  still incomplete in exploring magnetic frustration phenomenon which this crystal structure seems to facilitate for intersite antiferromagnetic interaction. We have therefore studied polycrystals of Tb and Er members  by various bulk experimental methods, including low-field magnetic  susceptibility ($\chi$), ac $\chi$ and isothermal remnant magnetization ($M_{IRM}$)  measurements which Gorbunov et al [11] did not carry out on the single crystals of Tb compound to search for glassy features. The physical properties of the Er member [4]  were not known before. We find that the  polycrystals of the Tb compound indeed reveal new features, including cluster spin-glass anomalies.

II.   Experimental details

Polycrystalline specimens of $Tb_3Ru_4Al_{12}$ and $Er_3Ru_4Al_{12}$ in single phase were synthesized by arc melting stoichiometric amounts of high purity constituent elements in an arc furnace in an atmosphere of argon, annealed at 800 °C for a week, and characterized  by x-ray diffraction (Panalytical XPert MPD, Cu K$_\alpha$) and scanning electron microscope (SEM) (Ultra Field Emission SEM of Zeiss). X-ray diffraction patterns are shown in figure 1c and figure 1d for both the compounds, including the results of Rietveld fitting. The lattice constants obtained are in good agreement with those reported in the literature. Temperature (*T*) dependencies  (2 – 300 K) of dc and ac magnetization (*M*) on a rectangular piece of about 18 mg were performed with a commercial SQUID magnetometer (Quantum Design, USA). Heat-capacity (*C*), electrical resistivity (ρ) and magnetoresistance [MR= [ρ(H)-ρ(0)]/ρ(0)] were probed with the help of a commercial Physical Properties Measurements System (Quantum Design, USA). For other experimental details, the readers may see Ref. 14.

## III. Results and discussion

### A. $Tb_3Ru_4Al_{12}$

*Dc magnetic susceptibility:*

Dc magnetic susceptibility, measured in a field of (H=) 5 kOe, is plotted in the forms of χ versus T and $χ^{-1}$ versus T in figure 2. The data shown in this figure is obtained after cooling the specimen from 300 to 2 K in zero field. χ exhibits Curie-Weiss behavior above 175 K. There is a deviation from the high-T Curie-Weiss behavior as the temperature is lowered well below 175 K, which is attributable to a gradual development of short-range magnetic correlations and/or crystal-field effects with decreasing temperature [11]. The effective moment ($μ_{eff}$) obtained from the high temperature Curie-Weiss region is found to be 10.2 ± 0.05 $μ_B$/Tb ion; this value is closer to the theoretical value of 9.72 $μ_B$ (for trivalent Tb) than that reported on single crystals (10.4 to 10.7 $μ_B$/Tb ion for different orientations) [11]. The value of paramagnetic Curie temperature ($θ_p$) turns out to be 65 K, with the positive sign signaling that the exchange interaction is of a ferromagnetic type for the fully degenerate trivalent Tb ions in this compound. However, in the low-temperature χ data, the presence of a peak at 22 K suggests onset antiferromagnetic ordering, instead of ferromagnetic ordering. One possible explanation for this is that intralayer exchange interaction is ferromagnetic, whereas interlayer interaction is antiferromagnetic with a net antiferromagnetism in the magnetically ordered state. It is not uncommon to see such antiferromagnetic and ferromagnetic exchange interactions in layered compounds. Alternatively, the exchange interaction for the ground state following crystal-field-splitting could be of an antiferromagnetic type which naturally facilitates magnetic frustration in kagome lattices.

In the plot of χ versus T (see figure 2), following a peak at 22 K with decreasing temperature, there is another upturn below about 20 K. In order to understand this better, we have measured χ with 100 Oe with increasing temperature for two protocols: zero-field-cooled (ZFC) and field-cooled (FC) from about 200 K (that is, from the paramagnetic state) to 1.8 K. The shapes of the curves obtained (with the values in arbitrary units) are shown in figure 3a in the low temperature range. It is clear from this figure that, in addition to the sudden increase in slope around 22 K with decreasing T, marking the onset of long-range magnetic order, there is a rounded peak at 12.3 K in the ZFC branch and the FC curve bifurcates in the vicinity of this temperature. The bifurcation of these curves indicates possible existence of a spin-glass phase near 12 K, as though this compound could be characterized as a re-entrant spin-glass. A noteworthy point is that the irreversibility of ZFC-FC curves sets in at a temperature ($T_{irr}$= 16.4 K) which is higher than the peak temperature ($T_p$= 12.3 K) in ZFC curve (see figure 3a) unlike in canonical spin-glasses (15) for which $T_{irr} ≤ T_p$; in addition, the FC curve continues to increase below $T_p$, whereas, for canonical spin-glasses, FC curve should remain flat. Such observations are usually attributed to random freezing of magnetic clusters. For such reports on cluster spin-glass behavior, the readers may see the articles, for instance, on doped $CaRuO_3$ [16], $La_{1.85}Sr_{0.15}CuO_4$ [17], $CeNi_{1-x}Cu_x$ [18], and $Li_3Ni_2RuO_6$ [19]. Even in stoichiometric compounds, triangular arrangement of antiferromagnetically coupled magnetic ions has been known to result in such cluster-glass features, e.g., $U_2IrSi_3$ [Ref. 20].

We get another new information based on dc χ behavior measured with 50 Oe for the field-cooled-cooling (FCC) and field-cooled warming (FCW) condition. The curves obtained are shown in figure 3b. The behavior is hysteretic in the range 12 – 20 K. The exact origin of this is not clear at this moment. A question arises, whether there is a subtle first-order transition setting in below 22 K, however broadened by crystallographic disorder. [Hysteretic nature of the magnetic transition is supported by ρ data as well (see below)].

*Isothermal magnetization:*

It may be recalled that sharp changes in *M(H)* at 2 K for the [001] orientation of the crystal were reported in Ref. 11. We show our data below 40 kOe on the polycrystals in figure 4 for various temperatures, 2, 8 and 15 K. It is obvious that a jump appears in each of the *M*(*H*) plots at low fields (~1 kOe) in the magnetically ordered state as in single crystals. This feature is usually attributed to the response of the magnetic clusters to the initial applications of magnetic fields; that is, each cluster is at first saturated along their local easy axis. Another upturn appears around 12 kOe. *M(H)* curve at 2 K is hysteretic. The observed magnetic-field induced jumps at 2 K is characteristic of (disorder-broadened) first-order type transitions, as the virgin curve lies outside the envelope curve [21-23]. The area of the hysteresis loop decreases with increasing temperature. Finally, there is another upturn near near 70 kOe (see the inset of figure 4) and this is consistent with the presence of an additional metamagnetic transition at higher fields, as noted in single crystals [11]. It is not clear at present whether the metamagnetic transitions are caused by crystal-field level crossings induced by Zeeman energy. As inferred from Ref. 11, it appears that *M* continues to increase gradually even beyond 70 kOe and very high fields (~140 kOe) are required to attain the free ion value. Such a gradual variation of isothermal *M* with *H* is also a signature [18, 20] of cluster spin-glass.

*Isothermal remnant magnetization:*

Superparamagnetic systems and conventional ferromagnets under certain circumstances also show irreversibility in ZFC-FC curves. Therefore, there is a need to search for more features characterizing spin-glasses. We have therefore obtained $M_{IRM}$ curves at different temperatures (1.8, 4, 8 and 15 K). For this purpose, we have zero-field-cooled the sample to the desired temperature, switched on a magnetic field of 5 kOe and switched off after 5 mins. It took about 150 s for the magnet to come to zero field. $M_{IRM}$ measurements were started at this time (*t*= 0) . There is a dramatic change in the behavior of $M_{IRM}$ as a function of *t* as the temperature is varied. The value of $M_{IRM}$ after switching of the field, that is, at *t*= 0, for T≤8K, is significant (~ 6.16, 1.8, and 2 emu/g) and decays slowly with *t* (see figure 3c). This supports the proposal of glassy state of magnetism at these temperatures. However, for *T*= 15 K, the value of $M_{IRM}$ at *t*= 0 is significantly small (~ 0.6 emu/g) compared to that for 8 K, remaining almost flat thereafter with varying time. At 20 K, the values fell to an insignificant value and the decay could not be observed. The behavior at 15 and 20 K establish that spin-glass behavior is lost at these temperatures. The curves below $T_N$ were fitted to a stretched exponential form usually encountered [24] in cluster spin-glass systems, $M_{IRM}(t) = M_{IRM}(0) + A \exp(-t/\tau)^{1-n}$. The constant A arises from glassy component mainly contributing to observed relaxation effects. The time constant τ and the exponent n are related to the relaxation rate of the cluster spin-glass phase at respective temperatures. The large magnitude of τ (~1350 and 1200s at 1.8 and 8 K respectively) suggests stiffening [25, 26] of the relaxation, compatible with increased size, consistent with clustering nature of the spin-glass component. Though theoretically, the value of the exponent, 1-n, can fall in the range 1/3 to 1 for spin-glasses [27], a value of ~ 0.5 is indicative [25, 26] of the distribution of relaxation times. Thus, these parameters speak in favour of cluster spin-glass behavior [see also, Refs. 19, 28, 29].

*Ac susceptibility:*

To further characterize spin-glass behavior, we have measured ac susceptibility (ac field = 1 Oe) with various frequencies (ν). The results are shown in figure 5. The real part (χ′) of ac χ exhibits a peak at all frequencies and the curve shifts to a marginally higher temperature with ν (from 15.5 K for 1.3 Hz to 16.3 K for 1333 Hz). Similar peak is seen in the imaginary part (χ″) as well below 20 K. But, the value of χ″ becomes zero at $T_N$ only. Considering that $M_{IRM}$ behavior is not supportive spin-glass freezing above 15 K, this observation is interesting and we do not fully understand this χ″ behavior at present. Possibly, disorder-broadened first-order transition occurring in the range 12 – 20 K (as a function of *T*, proposed above) may result in a complex magnetic state (magnetic phase-co-existence), resulting in persistence of χ″ till $T_N$. Clearly, phase co-existence region (12-20 K) can behave differently from cluster spin-glass region (<12 K), as discussed for $Nd_7Rh_3$ [23]. In a field of 5 kOe, the χ′ curves obtained with different frequencies merge, resulting in a clear peak at the long range magnetic ordering temperature (22 K), and the χ″ peaks are completely suppressed. All these ac χ features support the existence of spin-glass freezing in zero field in the magnetically ordered state.

It is important to note that χ′ peak appears at a higher temperature (~15.5 K for 1.3Hz) compared to that in ZFC curve in dc χ, a behavior known for cluster spin-glasses [see, for instance, Refs 17, 19]. In order to render further support for the interpretation in terms of cluster spin-glass freezing, we fitted (see the inset of figure 5) the peak temperature in χ′ (denoted here by $T_f$) using the empirical Vogel-Fulcher law [15] given by ν= $ν_0$ exp [-$E_a$/$k_B$($T_f$-$T_o$)]. Here $T_o$ is the Vogel-Fulcher temperature, which is a measure of the intercluster interaction strength, $k_B$ is the Boltzmann constant, $E_a$ is the activation energy and $ν_o$ is the characteristic frequency inversely related to the relaxation time of the magnetization vector near $T_f$. In cluster glass systems, $ν_o$ covers a wide range of values from $10^6$ to $10^{13}$ Hz [30] and the small value obtained by fitting (of the order of $10^8$ Hz) in our case is consistent with the interacting magnetic clusters. The observation of non-zero value of $T_0$ (~14.5 K) also implies that the clusters interact with each other.

The peak in the imaginary part (χ″) appears at a lower temperature (~ 7K) for 1.3 Hz; similar observations were made for another cluster spin-glass system, $Li_3Ni_2RuO_6$ [19]. However, there is a change of slope around 15.5 K, as marked by a vertical arrow in Fig. 5b. Appearance of such multiple features in χ″ is attributable to the existence of a distribution in freezing temperatures of various clusters. Clearly, the clusters respond differently to measurements of real and imaginary ac χ, as one enters the magnetically ordered state. Possibly, such a kind of cluster-formation is presumably triggered by some degree of crystallographic defects/disorder/grain-boundaries intrinsic to polycrystalline form. Isothermal *M* behavior presented above (that is broadening of the first-order transition and virgin-curve behavior) is supportive of the existence of disorder.

We note that the χ′ curves obtained with a dc field of 5 kOe tend to deviate near 35 K from the zero-field curves. We attribute it to the dominance of short-range correlations well above $T_N$. Such a magnetic precursor effect [31] seems to be the hallmark of this family of compounds, as observed for Gd and Dy compounds [12, 14]. This finding, viewed together with cluster spin-glass behavior support the idea [32] that various collective magnetic interactions compete in the event of geometrical frustration.

*Heat-capacity:*

We show *T*-dependence of heat-capacity in figure 6. There is a distinct λ-type anomaly near 22 K, establishing the onset of long range magnetic order. The peak temperature gets lowered marginally for an application of a magnetic field, say 10 kOe, and smeared for 30 kOe (see top inset of figure 6). This downward shift of the peak is consistent with the antiferromagnetic nature of the 22 K transition. It is important to note that there is no other peak at lower temperatures and *C* monotonically decreases with *T* below the peak temperature. This is consistent with the glassy nature of any magnetic transition well below $T_N$. We have also derived the 4f-contribution ($C_m$) to heat-capacity as described in Ref. 14. For this purpose, we employed the *C(T)* for the Y analogue, $Y_3Ru_4Al_{12}$, reported by us earlier in Ref. 14. The value of γ for the Y compound obtained from the linear region in the plot of *C/T* versus $T^2$ was found to be ~38 mJ/mol K2; the value of $θ_D$ is ~300 K. Following the procedure suggested by Blanco et al [33] to account for the mass differences between Y and Dy compounds, we derived the phonon part for the Tb compound. The magnetic contribution ($C_m$) to heat capacity obtained by subtracting the phonon part is plotted in figure 6. The magnetic entropy ($S_m$) derived from the $C_m$ curves is plotted in the bottom inset of figure 6. The value of $S_m$ at $T_N$ turns out to be about 7.5 J/ Tb mol K, which is far less than that expected for fully degenerate trivalent Tb ion. That is, for J= 6 of Hund's rule ground state of $Tb^{3+}$, the value magnetic entropy should be equal to *R*ln13 (21.34 J/mol K). Therefore, crystal-field effects should be responsible for this reduced value of $S_m$ at $T_N$.

*Electrical resistivity:*

Temperature dependence of electrical resistivity while warming is shown in figure 7. The zero-field behavior, reported by Gorbunov et al [11] on single crystals, is in qualitative agreement with that seen in our data. The temperature coefficient of ρ is positive well above $T_N$ and there is a weak upturn as $T_N$ is approached below about 30 K due to enhancement of critical scattering contribution. Following this, there is a drop due to the loss of spin-disorder contribution. In the presence of a magnetic field, say, for *H*= 10 kOe, the features are similar (not shown). However, for *H*= 30 kOe, the upturn due to critical scattering as $T_N$ is approached is suppressed and d*ρ*/d*T* remains positive in the entire *T*-range of investigation. The curves for *H*= 0 and 30 kOe overlap above about 50 K (indicating negligible magnetoresistance), but deviating at lower temperature. This may be corroborated to the deviation of zero-field and in-field ac χ curves above $T_N$. Clearly, this finding supports the existence of magnetic precursor effect.

We show the ρ curves obtained while cooling and warming the specimen in the inset of figure 7 in the *T*-interval 11 to 20 K. It is obvious that there is a hysteretic behavior supporting the conclusions from the low-field dc χ data, presented above.

*Magnetoresistance:*

We have obtained ρ behavior as a function of *H* at low temperatures. Let us look at the virgin curves first. MR, defined as [ρ(*H*)-ρ(0)]/ρ(0), is negative for virgin curves at all the temperatures, 2, 8, 15 and 25 K. It is obvious from figure 8 that the shape of the curve at 25 K looks qualitatively different from those for *T*<*T*$_N$ and is similar to that of paramagnets. As argued in Ref. 11, the negative sign of MR below 22 K can arise from closing of antiferromagnetic energy gap in some portions of the magnetic Brillouin-zone by the application of external magnetic fields. Looking at the virgin curve of 2 K, there is a sharp change of slope near 10-20 kOe beyond which there is a gradual increase in the magnitude of MR; subsequently, there is a peak near 62 kOe. These features are comparable with the MR behavior reported for different orientations of the single crystal [11]. Thus, the field-induced magnetic transitions for different orientations are sensed in the polycrystals as well. The studies on single crystals were reported for virgin curve only by Gorbunov et al [11]. But, we have measured the MR behavior for various field-cycling, that is, for the sequence $H = 0$ to 70 kOe to -70 kOe to 70 kOe to 0. As seen in figure 8, the virgin curve lies outside the envelope curve. This is a signature of disorder-broadened first-order magnetic transition induced by magnetic-field [22]. In such situations, the high-field state remains frozen even on withdrawal of the applied field and this phenomenon is called 'kinetic arrest' [21, 22]. Generally speaking, the ferromagnetically aligned high-field state is characterized by a relatively lower value of electrical resistivity. Therefore, on returning the field to zero value, it is expected that the value of $\rho$ for the kinetically arrested phase should be lower with respect to the virgin state value. The fact that the zero-field value after field-cycling is higher than that of the virgin state in the present compound implies that the 'arrested' magnetic state is characterized by a complex scattering process, yet to be understood. As the temperature is increased, say to 8 and 15 K, the drops/peaks due to metamagnetic transitions (seen for 2 K) and the virgin curve behavior (lying outside the envelope curve) persist. However, at these temperatures, virgin curve as $H \rightarrow 0$ lies above the envelope curve unlike the situation encountered at 2 K. The zero-field value after field-cycling is intermediate between that at the highest field and virgin state value. This implies that, at $H= 0$ after field-cycling, the magnetic state is a mixture of virgin state and high-field state (that is, partial arrest of high-field state) at these temperatures. Clearly, there are subtle changes in the magnetism of this compound and consequently in the scattering process with increasing temperature towards $T_N$. The present investigations bring out an interesting magnetic phase co-existence phenomenon following cycling the specimen across metamagnetic transition fields at very low temperatures.

### B. $Er_3Ru_4Al_{12}$

*Dc Magnetization:*

$\chi$ undergoes a monotonic increase with decreasing $T$ (figure 9a). Inverse susceptibility plot is linear over a wide temperature range (that is, Curie-Weiss behavior) and there is a gradual deviation from linearity below about 75 K due to possible role of short-range magnetic correlations and / or crystal-field effects. The value of $\mu_{eff}$ obtained from the Curie-Weiss fitting turns out to be 9.44$\mu_B$ per Er, in good agreement with the theoretical value of trivalent Er (9.59 $\mu_B$). The value of $\theta_p$ is about 8 K and the positive sign implies ferromagnetic exchange interaction between Er ions in the high temperature range, as in the case of other heavy rare-earth members of this series. At this juncture, it may be recalled [12, 14] that the high-temperature $\theta_p$ values for Gd, Tb and Dy members are 80, 65 and 20 K, and the values for Tb and Dy members are higher than the de Gennes scaled values (with respect to that of Gd member) of 53 and 36 K respectively. The de Gennes scaled value (for full degeneracy) for the Er member should be about 13 K, and thus the observed value of $\theta_p$ is lower in contrast to the trend observed for Tb and Dy members. It is of interest to focus future studies whether a difference in the higher-order (e.g., quadrupolar) interaction of these two members is responsible for this anomaly. It has been shown [34] quantitatively in terms of the $B^0_2$ term that the anisotropic 4f orbitals can enhance the magnetic ordering temperature for a Tb compound even beyond that of the Gd analogue.

Now, looking at the low-temperature behavior, there is no feature attributable to the onset of long-range magnetic ordering in the *T*-range of investigation. There is no evidence for the bifurcation of ZFC-FC curves obtained in a field of 100 Oe (see figure 9b). Isothermal *M* curve at 2 K (see the inset of figure 9) exhibits a steep rise till about 10 kOe, followed by a very weak variation till the highest measured field of 70 kOe without any feature attributable to metamagnetic transitions unlike in Gd [12] and Tb cases. The high-field magnetic moment (4 $\mu_B$ per Er) at 2 K is far less than that expected for fully degenerate Er ion (9$\mu_B$) thereby supporting the existence of crystal-field effects.

*Heat-capacity:*

Figure 10 shows heat-capacity data. There is no evidence for any λ-anomaly above 2 K in the *C(T)*, measured till 40 K. However, after a monotonic decrease down to about 5 K, *C* exhibits a weak upturn (see the mainframe of figure 10). This weak upturn disappears for an application of 20 kOe and *C* tends to drop with a change of slope below about 3 K. The temperature at which this slope change appears increases with *H*, say, to about 4 K for 30 kOe and 7 K for 50 kOe (though smoothened). The upturn below 3 K in zero-field and the upward shift of this temperature with *H* suggest that the compound is on the verge of ferromagnetic order at 2 K. We have derived $C_{lattice}$ employing the values of Y analogue, as in Ref. 14, and $C_m$, which are shown in the inset of figure 10. It appears that $C_m$ exhibits a peak around 30 K, an origin of which could lie crystal-field effects and/or short-range magnetic correlations.

*Electrical resistivity and magnetoresistance:*

Electrical resistivity behavior is shown in figure 11a. In this figure, we have normalized the data to 100 K value, as it was extremely difficult to get absolute values of resistivity due to highly porous nature of this sample. In the zero-field data, there is no drop due to the loss of spin-disorder contribution. Thus, this property also rules out the existence of long-range magnetic ordering above 2 K. However, in the presence of an external magnetic field, say, 30 kOe, the zero-field and in-field curves deviate gradually below about 30 K, resulting in a gradual increase of the magnitude of MR with decreasing temperature. This feature with the negative sign of MR is supportive of the gradual dominance [31] of short-range magnetic order (which is suppressed by *H*) as the sample is cooled towards 2 K. We have also obtained MR(*H*) curve at 2 K (see figure 11b) and it was found to be non-hysteretic. The sign of MR remains negative for all magnetic fields and temperatures. At 2 K, there is a sharp increase in the magnitude of MR with *H*, attaining ~ -8% at 25 kOe. The sharp change for an application of a small magnetic field mimics that expected for weak ferromagnetic metals [35]. A large magnitude of MR at similar fields was also shown to arise from interacting small magnetic clusters of less than 100 atoms per cluster in metallic nanostructures [36]. Therefore, the large magnitude in the present case is consistent with short-range correlations proposed above. Absence of quadratic variation of MR with *H*, expected for paramagnetism, can also signal that there are critical point fluctuations which are suppressed by an application of a small magnetic field. This observation appears to be consistent with our proposal above that this compound is on the verge of magnetic order at 2 K. Beyond about 25 kOe, there is a gradual decrease in the magnitude and we attribute the upturn of the curve at higher fields to competing positive (metallic) contribution from the well-known influence of Lorentz force on the conduction electrons. The sharpness of the drop in MR at low-fields still persists at a slightly higher temperatures, say at 6 K (not shown here), but smoothened without any significant change in absolute values. We also noted that the values get gradually reduced at further higher temperatures. Such a finding implies that there is a gradual enhancement of short-range correlations with decreasing temperature, which might eventually result in magnetic order below 2 K.

IV.   Conclusions

We have investigated the magnetic behavior of metallic kagome lattices, $Tb_3Ru_4Al_{12}$ and $Er_3Ru_4Al_{12}$, by bulk experimental methods. We emphasize the following findings:

*In the case of the Tb compound*, we make some new observations with respect to the previous report on single crystals [11]. In addition to long-range antiferromagnetic ordering at 22 K as known earlier [11], cluster spin-glass features are seen at lower temperatures (around 12 K) in ac susceptibility, low-field dc susceptibility and isothermal remnant magnetization. In that sense, this compound could be classified as a re-entrant cluster-spin-glass. This spin-glass component was missed out in the earlier dc $\chi$ study [11], possibly because the magnetic field employed to measure dc $\chi$ was so large (50 kOe) that it suppresses spin-glass freezing. In view of our results, it is of interest to reinvestigate single crystals.

The following question naturally arises. If one claims that the crystallographic disorder and/or grain boundaries alone could be the origin of spin-glass anomalies in the polycrystalline form of the Tb compound, why is that the polycrystalline Gd analogue does not exhibit [12] any cluster spin-glass feature? In this connection, it may be recalled that the polycrystals of Dy analogue also show glassy characteristics [14]. These observations can be consistently understood by proposing that higher order interactions (e.g., quadrupolar) due to nonspherical symmetry of the orbital (responsible for magnetic ordering) may play a crucial role to trigger glassy dynamics in such topologically frustrated systems. Recently, a theory has been developed [37], bringing out that geometrical frustration can introduce cluster freezing, even in the low levels of disorder (including grain boundaries) in kagome lattices. Our results add a new dimension to this proposal, that is, to work out a theory exploring the role of higher order interactions leading to geometrical frustration in kagome lattices.

In addition, the magnetic susceptibility and electrical resistivity data as a function of temperature presented in this article bring out hysteresis in the range 12 – 20 K, thereby suggesting that the onset of magnetic ordering is followed by a disorder-broadened first-order type. This is an intriguing finding, and the exact origin of such a possible first-order transition is not clear to us. Similar hysteretic nature is brought out for the magnetic-field induced transitions also by this study. Magnetoresistance loops reported in this article bring out interesting magnetic phase co-existence phenomenon, following cycling across metamagnetic transition fields.

*With respect to Er compound*, no magnetic investigation has been known in the past literature. We do not find evidence for long range ordering down to 2 K. The observed data indicate that this compound is on the verge of long-range magnetic order at 2 K in zero-field and ferromagnetic behavior close to 2 K can be induced by an application of a small magnetic field as inferred from heat-capacity behavior.

In conclusion, this work emphasizes that, in this kagome family of materials, an interplay between geometrical frustration, consequences of the asphericity of the orbitals and possibly crystallographic defects could be studied to understand various ground states. It appears that this interplay is manifested differently for different rare-earths in this family, apart from a common feature of short-range magnetic correlations persisting over a wide temperature range above respective $T_N$. For instance, in the case of the Gd compound, Griffiths-phase-like features were reported [12] above $T_N$, whereas for the Tb and Dy cases, cluster spin-glass anomalies are seen. Thus, as proposed earlier [12, 32], various collective magnetic states compete in a complex fashion as a result of this interplay. It is of interest to understand temperature dependent magnetic structure, short-range correlations and crystal-field effects in these compounds.

Figure 1:
Crystal structure of $R_3Ru_4Al_{12}$ (R= rare-earths) to reveal (a) unit cell, and (b) planar and distorted kagome net of $R_3Al_4$ layer viewed along c-axis. X-ray diffraction (Cu $K_\alpha$) patterns for (c) $Tb_3Ru_4Al_{12}$ and (d) $Er_3Ru_4Al_{12}$. Rietveld fittings, along with fitted parameters and lattice constants, are also included in (c) and (d).

Figure 2:
Temperature dependence of magnetic susceptibility and inverse susceptibility obtained in a field of 5 kOe for $Tb_3Ru_4Al_{12}$. The continuous line in the inverse $\chi$ plot above 175 K is obtained by Curie-Weiss fitting.

Figure 3:
(a) Magnetic susceptibility behavior below 40 K taken in a field of 100 Oe for $Tb_3Ru_4Al_{12}$ for the zero-field-cooled and field-cooled states of the specimen. (b) The curves obtained in a field of 50 Oe for the field-cooled-cooling and field-cooling condition while warming are plotted below 40 K to highlight hysteresis. The lines through the data points serve as a guide to the eyes. (c) Isothermal remnant magnetization behavior at 1.8, 4, 8 and 15 K; the lines through the data points are obtained by fitting to a stretched exponential form, mentioned in the text.

Figure 4:
Isothermal magnetic hysteresis loops (below $H$= 40 kOe) obtained at 2, 8 and 15 K for $Tb_3Ru_4Al_{12}$. The virgin curve up to 70 kOe for 2 K is shown in the inset. The numericals and arrows are drawn to guide the eyes.

Figure 5:
Real and imaginary parts of ac susceptibility as a function of temperature for $Tb_3Ru_4Al_{12}$, measured with various frequencies ($\nu$= 1.3, 13, 133 and 1333 Hz) in zero field and in 5 kOe. The lines through the data points serve as guides to the eyes. Tilted arrows are drawn to show the direction in which the curves move with increasing frequency. The inset shows a fit of the peak temperature in $\chi'$ as described in the text. In $\chi''(T)$ plot, a vertical arrow is drawn to show a slope change in low-frequency curves.

Figure 6:
(Mainframe) Temperature dependence of heat-capacity in the absence of any external magnetic field for $Tb_3Ru_4Al_{12}$. A line is drawn through the data points to serve as a guide to the eyes. The lattice contribution and the 4f contribution ($C_m$) to $C$ (obtained as described in the text) are also shown. The zero-field and in-field (10 and 30 kOe) curves below 40 K are shown in the top inset. In the bottom inset, magnetic entropy (in zero field) derived from the $C_m$ curve is plotted as a function of temperature below 60 K.

Figure 7:
Temperature (2 - 300 K) dependence of electrical resistivity for $Tb_3Ru_4Al_{12}$ in zero field and in 30 kOe. Inset shows the data in a narrow temperature range (11 – 25 K) for the cooling and warming cycles to show the hysteretic behavior.

Figure 8:
Magnetoresistance as a function of magnetic field (in the field range -70 to 70 kOe) for $Tb_3Ru_4Al_{12}$ at 2, 8, 15 and 25 K. The numericals and arrows are placed to follow the curves with the change of magnetic field.

Figure 9:
(a) Temperature dependence of magnetic susceptibility and inverse susceptibility obtained in a field of 5 kOe for $Er_3Ru_4Al_{12}$. The continuous line in the inverse $\chi$ plot is obtained by Curie-Weiss fitting. (b) The data obtained for zero-field cooling (points) and field-cooling conditions (line) of the specimen with a field of 100 Oe. Isothermal magnetization behavior (per formula unit) at 2 K is shown in the inset.

Figure 10:
Heat-capacity as a function of temperature for $Er_3Ru_4Al_{12}$ in the range 2 – 25 K in zero field and in the presence of 20, 30 and 50 kOe. In the inset, the data in a wider temperature range (2-45 K) is plotted, along with the lattice and magnetic contributions to heat-capacity.

Figure 11:
Normalised electrical resistivity as a function of temperature (2-100 K) for $Er_3Ru_4Al_{12}$ in zero field and in 30 kOe. (b) Isothermal magnetoresistance behavior at 2 K and a line is drawn through the data points.

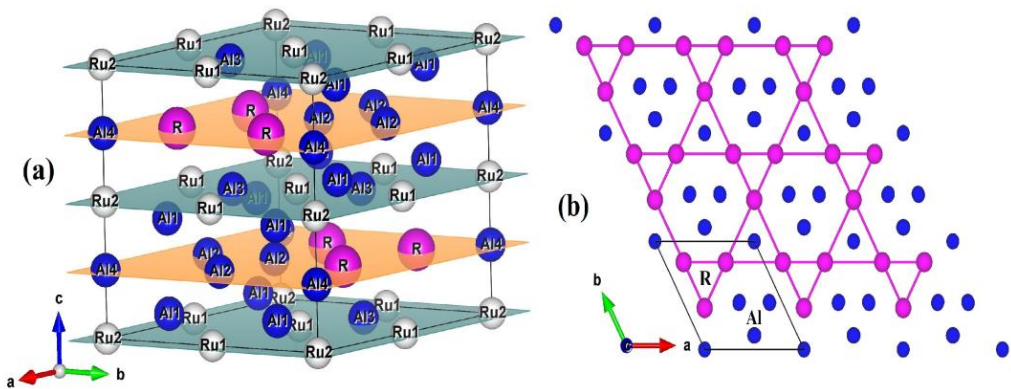
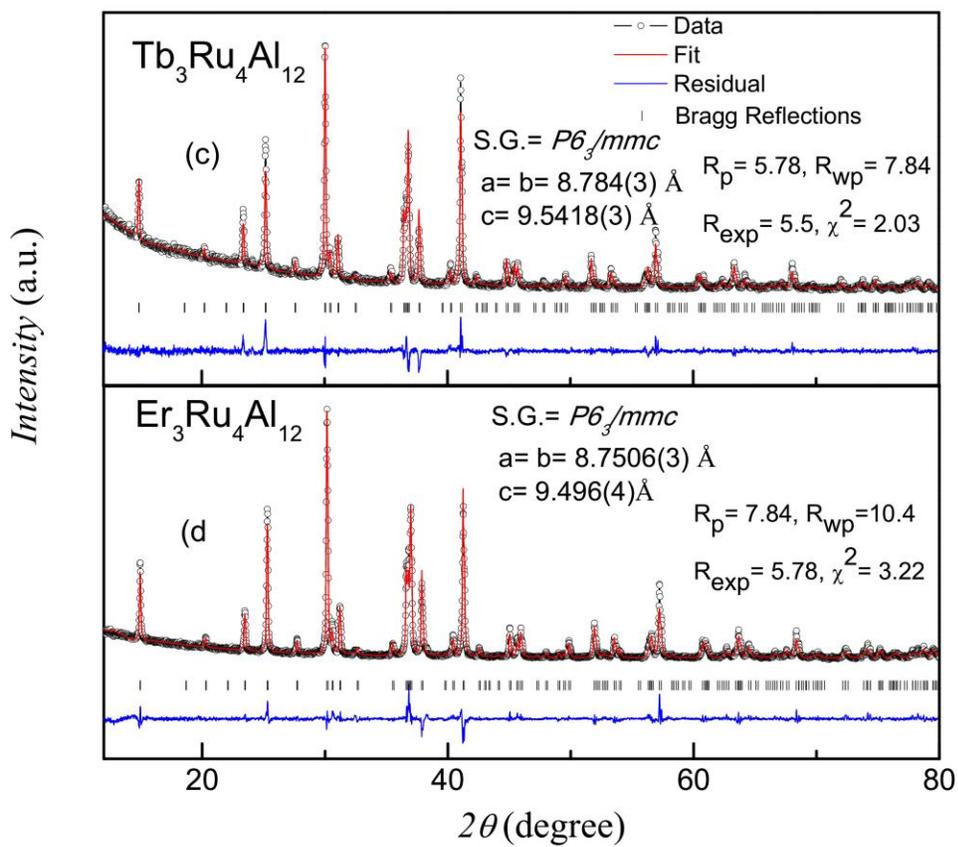

Figure 1

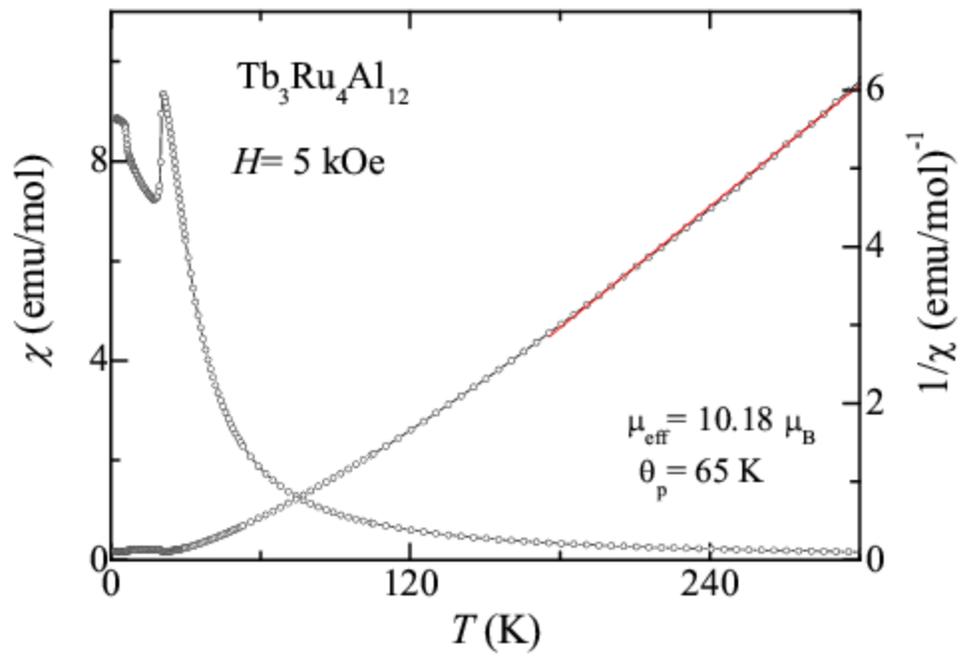

Figure 2

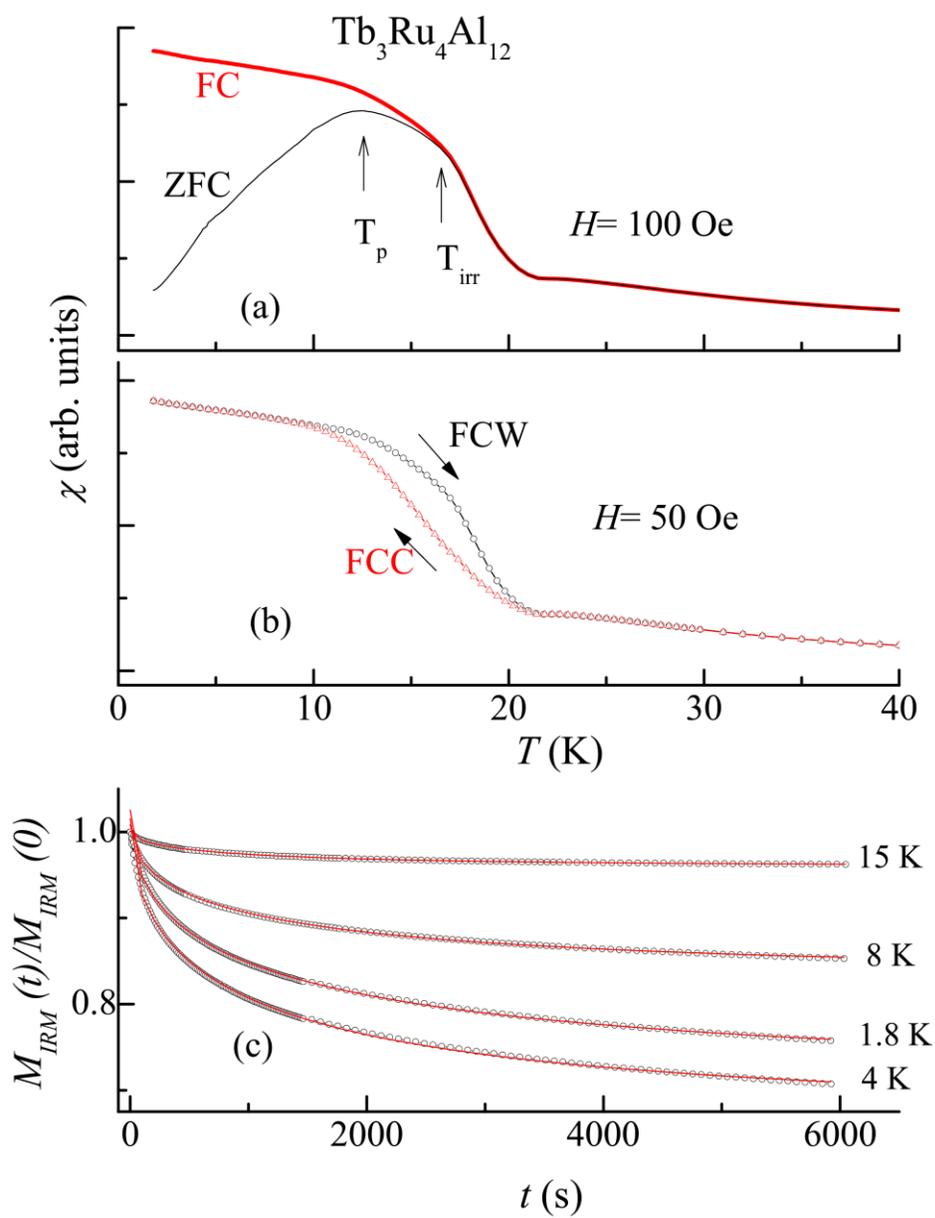

Figure 3

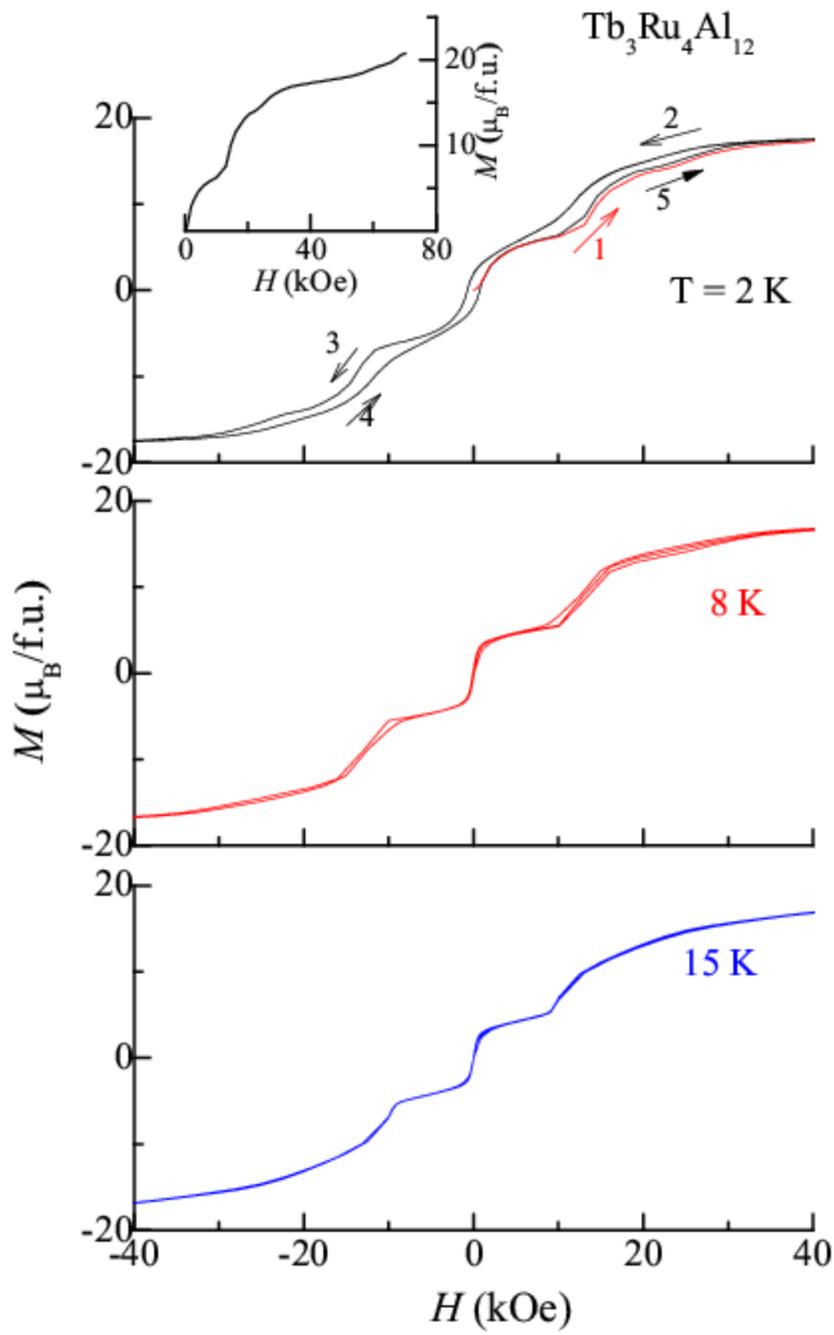

Figure 4

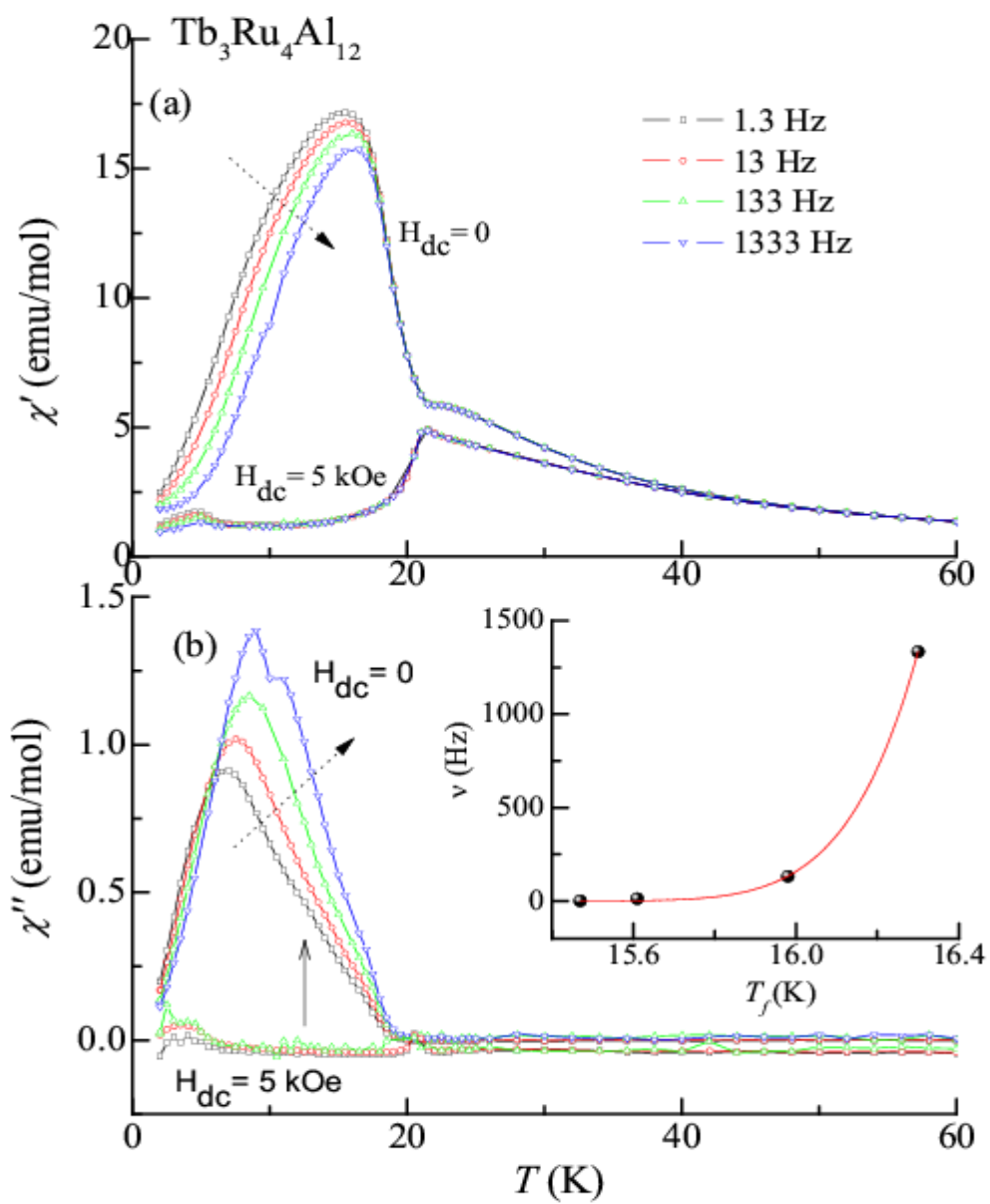

Figure 5

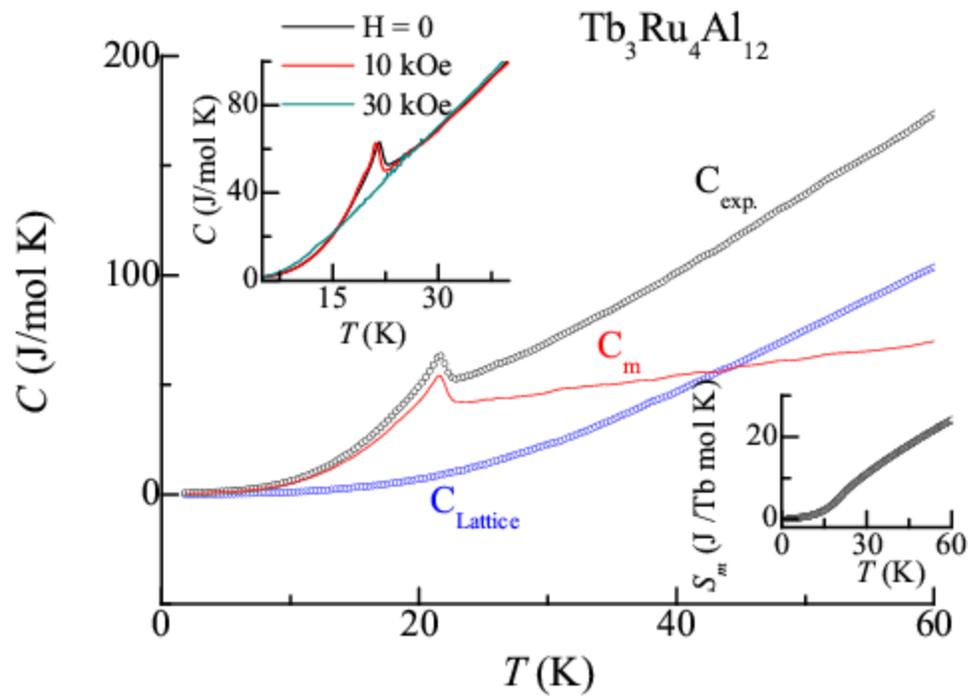

Figure 6

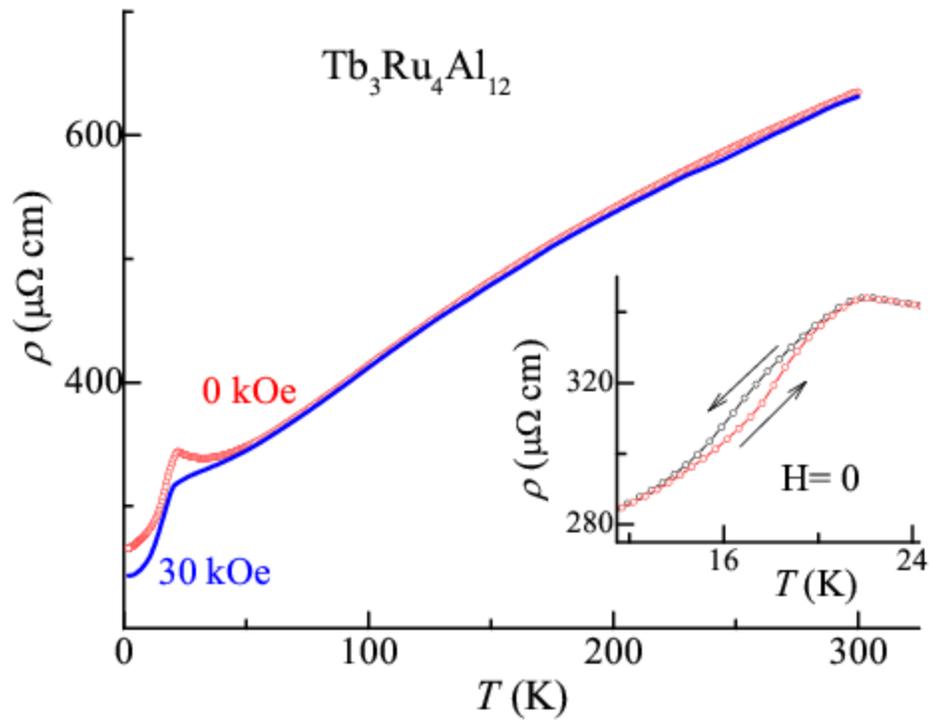

Figure 7

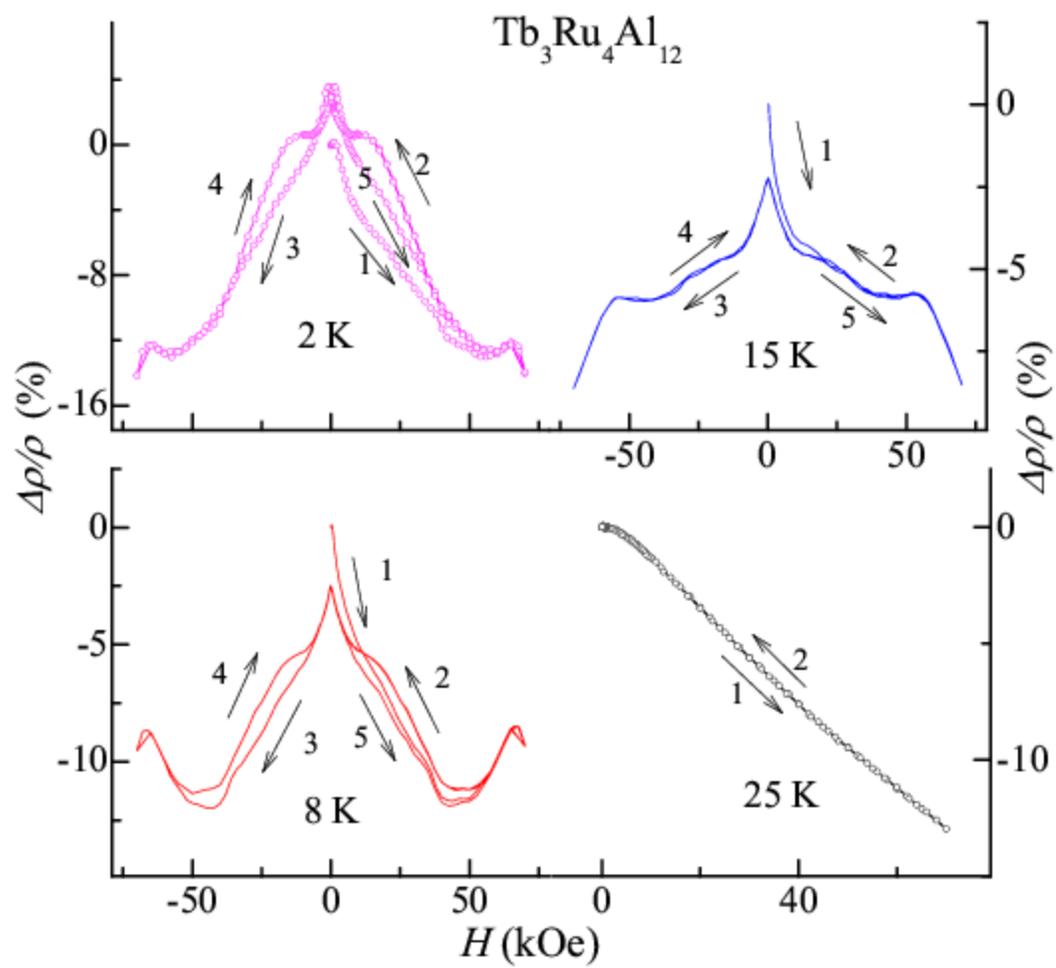

Figure 8

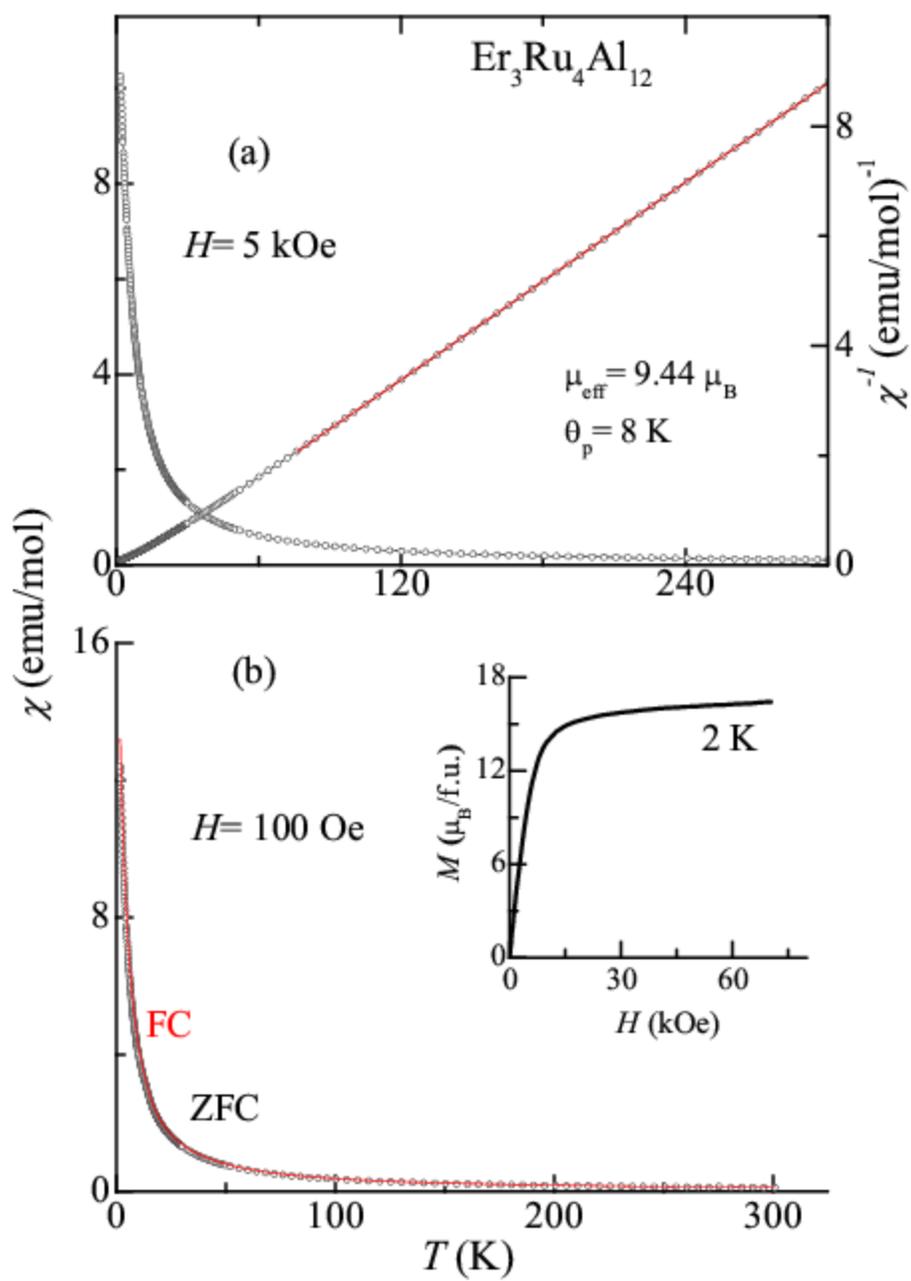

Figure 9

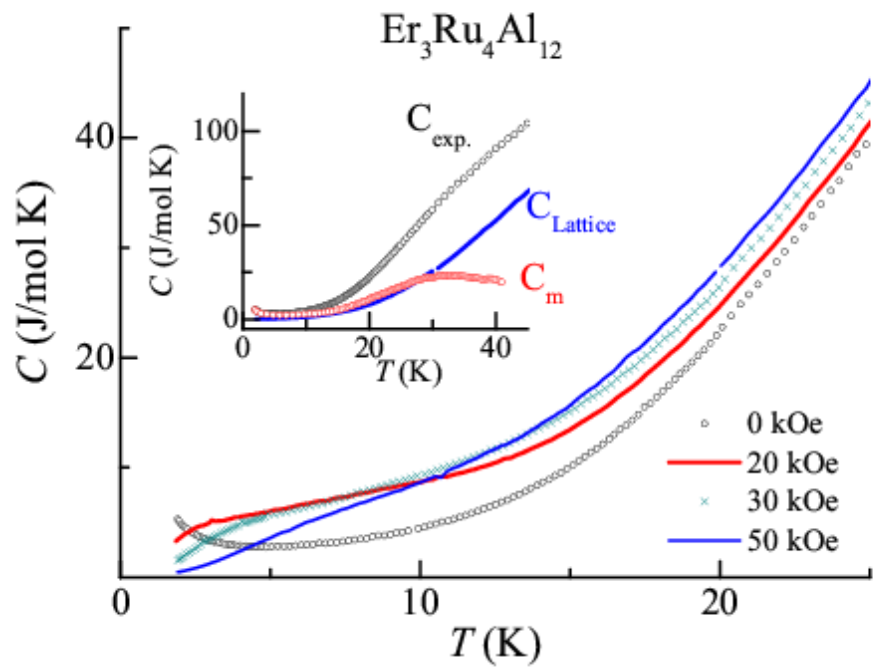

Figure 10

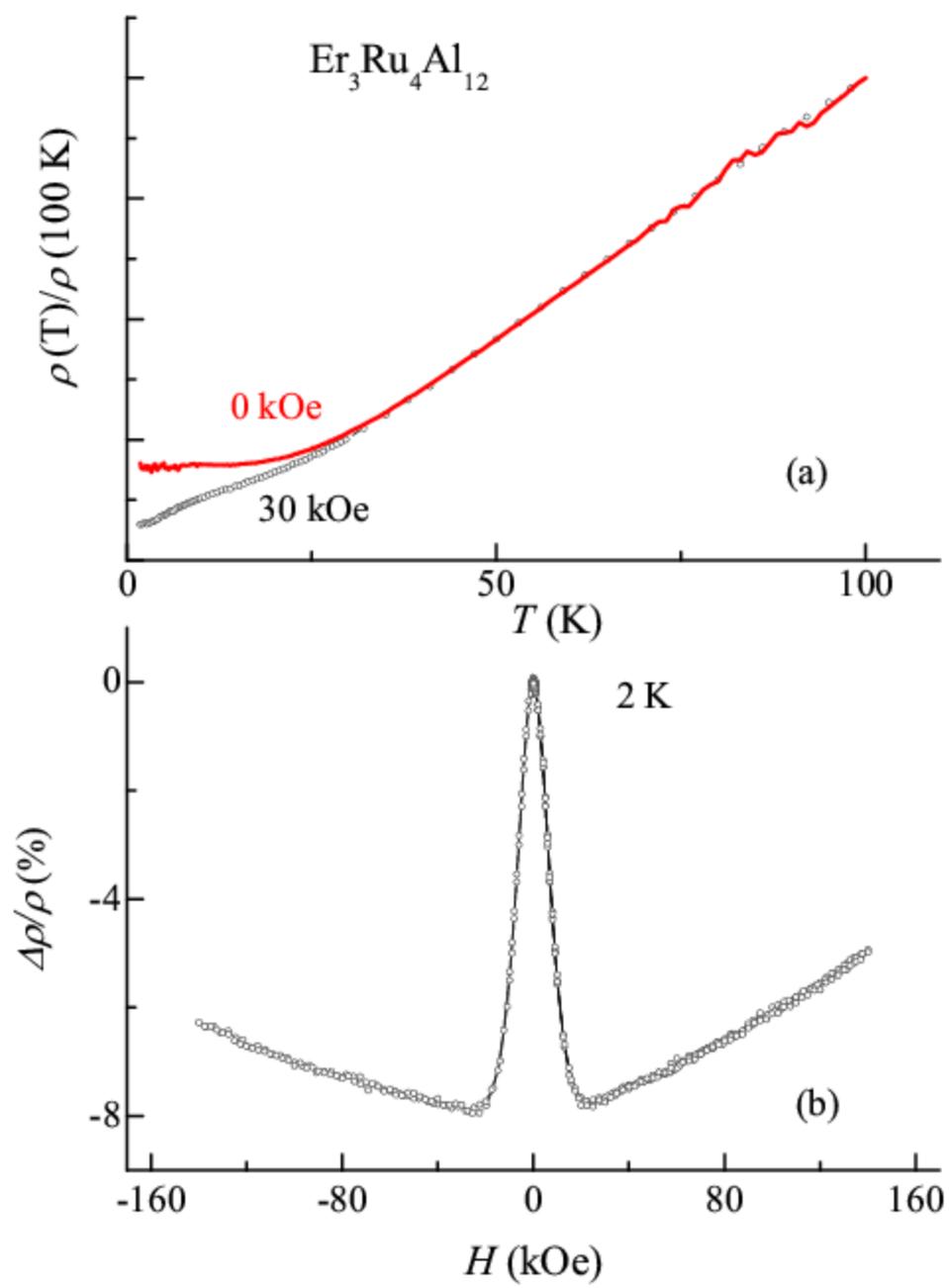

Figure 11